# Significant acceleration of development by automating quality assurance of a medical particle accelerator safety system using a formal language driven test stand

P. Fernandez Carmona, M. Eichin, A. Mayor, H. Regele, M. Grossmann, *Member, IEEE*, DC. Weber.

*Abstract*— At the Centre for Proton Therapy at the Paul Scherrer Institute cancer patients are treated with a fixed beamline and in two gantries for ocular and non-ocular malignancies, respectively. For the installation of a third gantry a new patient safety system (PaSS) was developed and is sequentially being rolled out to update the existing areas. The aim of PaSS is to interrupt the treatment whenever any sub-system detects a hazardous condition. To ensure correct treatment delivery, this system needs to be thoroughly tested as part of the regular quality assurance (QA) protocols as well as after any upgrade. In the legacy safety systems, unit testing required an extensive use of resources: two weeks of work per area in the laboratory in addition to QA beam time. In order to significantly reduce the time, an automated PaSS test stand for unit testing was developed based on a PXI chassis with virtually unlimited IOs that are synchronously stimulated or sampled at 1 MHz. It can emulate the rest of the facility using adapters to connect each type of interface. With it PaSS can be tested under arbitrary conditions. A VHDL-based formal language was developed to describe stimuli, expected behaviour and specific measurements, interpreted by a LabView runtime environment. This article describes the tools and methodology being applied for unit testing and QA release tests for the new PaSS. It shows how automation and formalization made possible an increase in test coverage while significantly cutting down the laboratory testing time and facility's beam usage.

## I. INTRODUCTION

PROTON therapy is a radiation therapy technique that uses protons accelerated to the hundred MeV range to deposit an ionizing dose meant to destroy the cancerous cells located in deep seated tumors. This therapy modality was first proposed by R. Wilson in his seminal paper published in 1946 [1]. The first patient treatments started in 1954 in Berkeley and at the Paul Scherrer Institut (PSI) it is in clinical use since 1984. At PSI patients are irradiated with a fixed beamline and in two gantries for ocular and non-ocular malignancies, respectively [2]. A third gantry is currently being commissioned to increase the capacity [3] [4]. They are named Optis and Gantry1 to Gantry3.

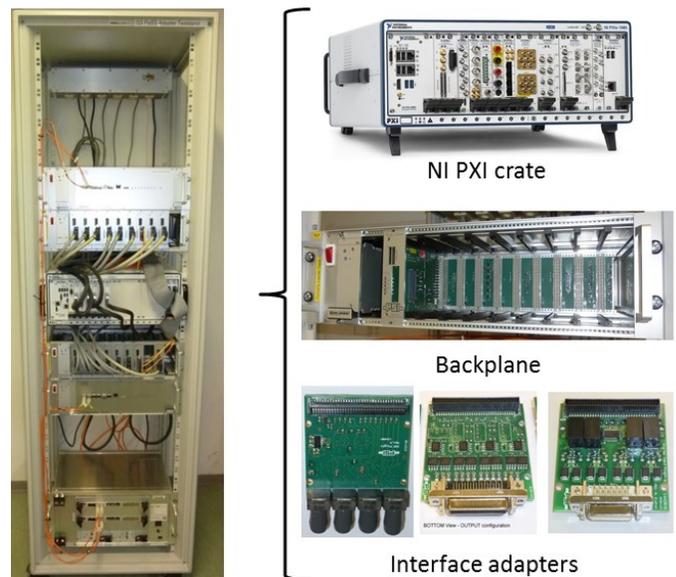

Fig. 1. Test stand hardware developed to automate the unit testing of the Patient Safety System

The main advantage of this treatment modality is the precise dose deposition of the protons at the end of its trail, in a narrow zone called the Bragg peak. This allows optimal sparing of healthy tissue in the vicinity of the target volume, therefore reducing potentially long term radiation-induced adverse events and secondary malignancies. One of the main disadvantages of proton therapy is the size and complexity of the accelerators and beam lines required to produce and transport, safely and precisely the particles to the patient.

At PSI the safety of the dose delivery is guaranteed by the Patients Safety System (PaSS). This is in place to monitor the facility and to interrupt the treatment whenever any sub-system detects a hazardous condition for the patient. For the installation of our third gantry, new PaSS technology was developed and is sequentially being rolled out to also update the existing areas [5]. To ensure correct treatment delivery, this system needs to be thoroughly tested during development,

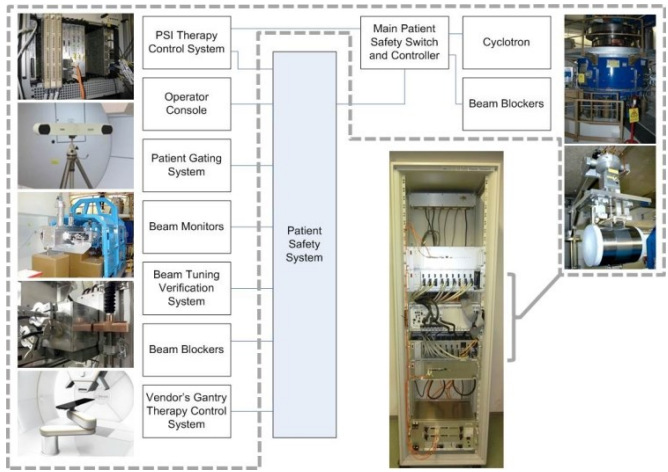

Fig. 2. Test stand for unit testing. It emulates every element of the facility connected to PaSS, such as sensors, actuators and control systems using a set of interface adaptors to each kind of signal.

as part of the regular quality assurance (QA) protocols and after any upgrade or modification.

In this article we will describe the design methodology used to guarantee a safe design and implementation of PaSS at our centre, and the tools recently developed to do it in a more efficient way: A test stand (Fig. **1**) was built to automatically run the laboratory unit testing, a language was developed to formally describe the individual unit tests and part of the periodic facility QA was automated. Finally the experience after two years and two treatment areas using this new technology will be presented.

## II. DEVELOPMENT AND QUALITY ASSURANCE FLOW

The central document describing the safety strategy at the Center for Proton Therapy at PSI is the "Report on Safety Measures". It includes the risk analysis, actions, severity levels and methods to be used in the facility to guarantee a safe treatment to the patients. The content of this document has been reviewed and agreed with the Swiss radiation protection competent authorities. The PaSS specification for each of the treatment areas is derived from the report on safety measures. The design specifications detail the functionality, internal structure and list of all hardware lines to be realized. Once the PaSS is implemented, it needs to be thoroughly tested before it is deployed for patient treatment with real beam. At PSI we have defined three sets of QA tests for PaSS:

Unit testing is performed on the PaSS logic hardware in a laboratory, emulating the inputs and evaluating the outputs for correctness without any real sensor or actuator connected. A set of unit tests are specified deriving from the design specification. Each test verifies the correct implementation of a specified function. In the last PaSS implemented there were over 400 individual unit tests executed.

System testing is performed on the facility by forcing different controls, sensors and accelerator parts into known states and evaluating the proper behavior of PaSS and the connected final elements, such as magnets, beam blockers or high frequency generators. The system tests are also derived from the design specifications and aim to verify the complete action change, from sensor to actuator. Being they performed on the real facility, it is not possible to test many combinations of states. Therefore a thorough prior unit testing in the laboratory is of the highest importance to guarantee safety.

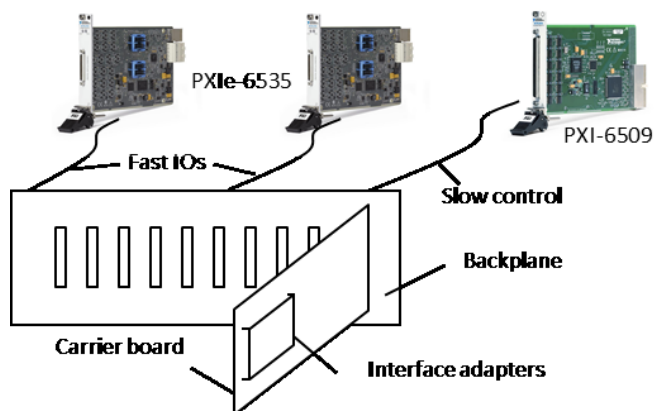

Fig. 3. Backplane, carrier boards and plugins to adapt different signal types that need to interface to PaSS with the 5V IOs of the PXI data cards.

Finally, regular facility QA tests are performed to guarantee the performance of PaSS and the final safety elements. There are daily basic tests performed together with the medical physics dosimetric QA every morning, before patients start being treated. More complete weekly, monthly and yearly tests are performed by the staff of the center, and operations are stopped should any test fail, until the problem is identified and solved.

## III. UNIT TESTING

The unit testing is performed in the laboratory on the PaSS hardware running the final safety logic, under controlled conditions. All input signals are driven to well defined patterns and all outputs are monitored to verify the correct response to stimuli. In the past the unit testing was performed using National Instruments CompactRIO™ in a modular manner. With a logic analyzer the different logic functions were sequentially tested connecting only the signals involved, due to the limitation of available channels. The unit tests were specified textually in documents including waveforms describing the stimuli and tables indicating the expected outputs. These specifications were manually introduced in a graphical tool that generated the stimuli patterns. With the occasion of the new Gantry 3 installation we developed a new unit testing platform and methodology.

### A. Test stand

The new Test stand is a rack mounted computer that can automatically execute unit tests by driving and monitoring multiple hardware lines connected to all PaSS IOs, see Fig. **2**. It is based on a National Instruments™ crate with up to eighteen PXI cards. The signal lines are driven using fast PXIe-6535 cards clocked at 1MHz. Slower PXI-6509 cards were used for control tasks, like defining the direction of lines and reading temperature sensors.

As shown in Fig. **3**, the routing of the PXI cards IO pins to the corresponding PaSS signals is achieved with an in-house designed 19 inch backplane board with 90° connectors to host carrier boards. On the carriers, different mounted plug-in interface adapters convert the 5V digital signals to and from the PXI crate into each corresponding interface type, such as optical, 24V digital lines and three wire, redundant current loops. Other plugin types could be developed in the future to interface other signal types.

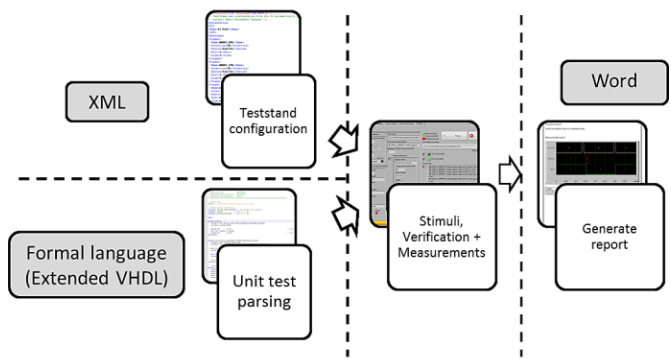

Fig. 4. Modular software structure of the Test stand application, with the different file formats currently in use.

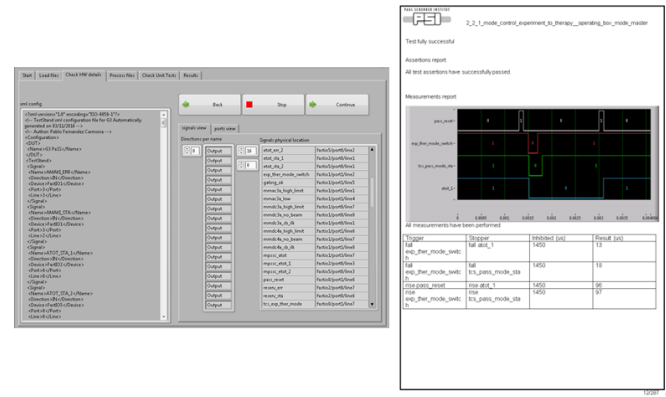

Fig. 5. Labview written test stand graphical user interface and a section of the generated unit testing report.

Each of the PXI crate used is limited to 400 IO lines, which was enough for testing both Gantry 3 and Optis safety systems. It is planned to update more PaSS areas with a higher signal count and the test stand already prepared for that, as the PXI crates allow for a daisy chain synchronization of several crates, therefore extending the available IOs to whatever might be necessary.

Running on the PXI crate there is a software application programmed in LabView. It is modular and extendable, as shown in Fig. 4, containing the following parts: the test stand configuration module, unit test parsing, test execution, results verification and reporting. The test stand configuration module parses an xml file describing the name and type of each of the signals and configures accordingly the fast IO lines, backplane and the connected plugins. The xml file is generated automatically with a tool provided by the PaSS framework that was developed for the integration of Gantry 3.

The unit test parser reads the test descriptions, performs a syntax check and creates an internal executable data structure containing stimuli, assertions and time measurements. It fetches any extra included files, expands macros and unrolls loops into multiple simple tests to be executed sequentially. The code is interpreted, not compiled.

The execution module uses the created internal data structure to drive the input signals to the specified stimuli patterns and monitors the status of the outputs. IO sampling and driving is done synchronously with a 1 µs time base. It also performs the programmed time measurements and verifies if all the specified assertions are met.

The report module analyses the internal execution data structure and creates a final document with one section per unit test executed. It generates a summary with a list of all tested items and its success, information of failing assertions, detailed tables and time diagrams of all the time measurement performed to facilitate the interpretation. Currently the report is written in MS Doc format based on a provided template file.

The user interface is a simple state machine with tabs to guide the user in loading the different configuration and test files, gives feedback of the internal structures created, syntax errors or hardware and configuration issues. After all the required files are loaded and the hardware configuration reviewed and approved by the user, all tests are autonomously executed. Finally the report is generated, ready to be checked and signed by the unit tester. See Fig. 5.

As the application is modular, with clearly defined interfaces and internal data structures, it would be possible to easily extend it to support new unit test description languages or to generate different report formats.

Some preliminary results of the design and development of the PaSS test steand were presented at [6].

*B. System calibration and validation*

The input and output signals of the test stand are timed using a shared hardware clock routed through the PXI bus and should therefore be very precise and stable. However, in order to guarantee the correct behavior of the test stand and to be able to trust the unit testing reports, it was calibrated.

In order to have an independent measurement we added a sniffing debug port to the carrier boards with direct access to the digital signals. To this debug port we connected a logic analyzer for crosscheck.

With the described setup we chose a subset of the real Gantry 3 unit tests, including one per PaSS logic function. These tests were manipulated in a way that the test stand application should detect errors at an expected time.

The tests were executed to generate a test report. It was then verified that all the tests failed as intended. Also, by comparing the generated waveforms from the report and the logic analyzer, as seen in Fig. 6 and Table I, it was confirmed that the time measurements matched.

A validation report was created and made available to the authorities as part of the release documentation of the new PaSS. Once validated, the test stand unit test reports were trusted in future iterations of Gantry 3 and Optis development.

*C. Tests description formal language*

After an investigation in literature of different existing languages to describe tests and assertions, nothing was found that was both compact and close enough to natural language as to be able to replace the textual description in the unit test documentation. The best option for our needs was VHDL syntax, which was extended with three new functions: Macros, expansion loops and time measurements. The code is used to describe three groups of functions, as seen in Fig. 7: stimulate inputs, define assertions to verify outputs and program time measurements.

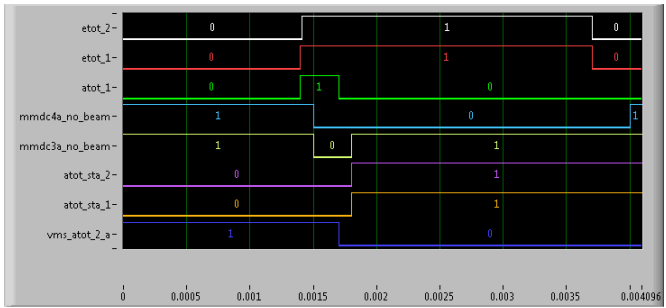

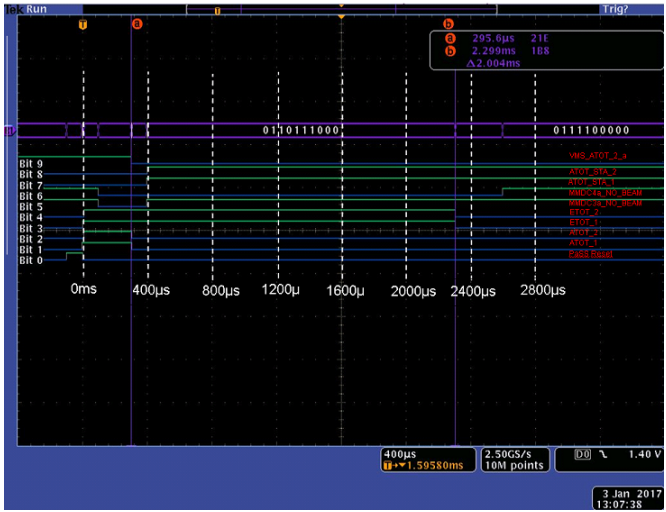

Fig. 6. Waveform generated by the test stand (above) and by the logic analyzer (below).

The unit tests can be implemented in a single or multiple files that can be chained with the include keyword. Macros are pieces of code that replace a placeholder when executed by the interpreter. The macro syntax includes a definition with a name and content, and a call:

```
DefineMacro MY_MACRO_NAME
Content of the macro;
EndMacro
…
callMacro MY_MACRO_NAME
```

Our main use of macros is to define typical time constants, initialization of a large number of signals and functions that will be reused and called from several unit tests.

Expansion loops are a custom extension used to repeat a same unit test under different conditions or variations. The syntax includes a keyword to open and close a loop, and tags that indicate how many different variables of this loop will be executed. The resulting amount of executed unit tests will be the product of the number of different tags of all the loops defined. The syntax is as follows:

```
Loop
  Tag Tag_name_1
```

Table I

TIME MEASUREMENTS PERFORMED

| Trigger | Stopper | Result (us) |
|---|---|---|
| fall vms_atot_2_a | rise atot_sta_1 | 100 |
| fall vms_atot_2_a | rise atot_sta_2 | 100 |
| fall vms_atot_2_a | rise mmdc3a_no_beam | 100 |
| fall vms_atot_2_a | rise mmdc4a_no_beam | 2300 |
| fall vms_atot_2_a | fall atot_1 | 6 |
| fall vms_atot_2_a | fall etot_1 | 2006 |
| fall vms_atot_2_a | fall etot_2 | 2006 |

Time measurements table extracted from the unit test report used for the test stand validation, corresponding to waveforms above in Fig. 6.

Fig. 7. The unit test language can describe three different aspects: stimuli, assertions and time measurements.

```
    Code option 1
    EndTag
    Tag Tag_name_2
    Code option 2
    EndTag
EndLoop
```

The result will be two independent tests, one including only option 1 code and the other one including only option 2. Both tests will share the code before and after the loop definition. A typical effort saving case would be to check a certain function under experimental and therapy modes, in combination with the allocation or non-allocation of mastership, therefore requiring two loops with two tags each.

As in VHDL, subsequent value assigns, or definitions of assertions of the same signal at the same delta time are overwriten. Only the last entry before a time advance will be active. This is useful when using general purpose macros that can be overwritten for single cases, for example inside a loop.

The final nonstandard extension introduced is time measurements. This allows a precise detection of the time span between two defined events, which can be rising or falling edges. The syntax is as follows:

```
measure [rising_edge/ falling_edge]
(TRIGGER_SIGNAL) to [rising_edge/ falling_edge]
(STOPPER_SIGNAL) name "Unique_name";
```

As discussed before, this language is interpreted and time driven. Signals can be assigned at any time, but only constant values; assigning the value of one signal to another one is not allowed.

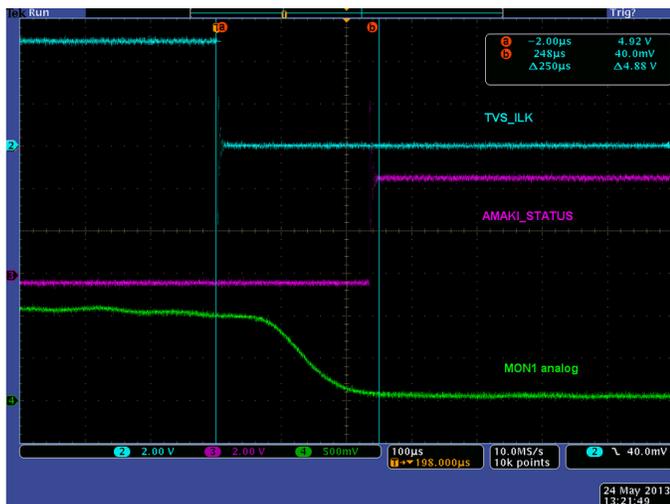

Fig. 9. The built in time measurements implemented in the new PaSS technology replace regular oscilloscope measurements like this one of the reaction time of a safety element and beam monitor after an interlock event.

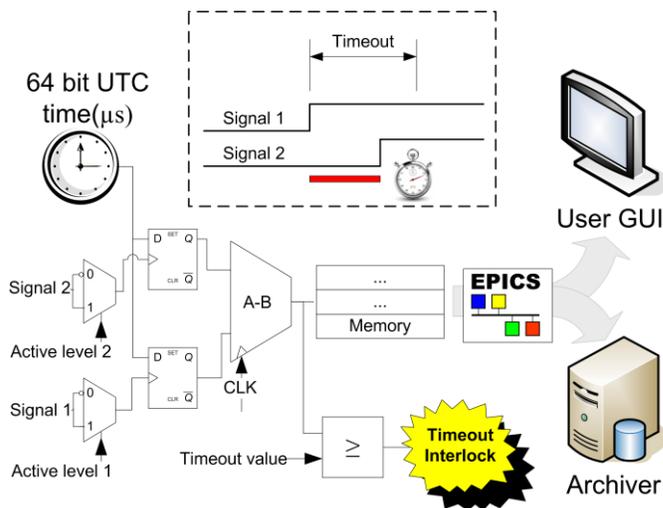

Fig. 10. Time measurements functionality built into the safety logic. The PaSS will trigger an interlock when certain reaction takes longer than a defined timeout value. The last measurement event is made visible via EPICS publication to the Graphical User Interface and archived for future analysis.

## IV. PARTIAL AUTOMATIZATION OF QA BY SELF-TESTING

Several yearly quality assurance tests of the facility involve reaction time measurements. Those are typically closing times of safety elements such as magnets or mechanical beam blockers and reaction times between an interlock occurrence and beam switching off at different monitors. Installing cables, an oscilloscope and performing these QA tests (Fig. 9) require one to two days of work a year per area. The tests also block the facility to be used for research, as they are typically performed in the evenings after patient treatment has concluded.

With the release of the new PaSS technology several self-testing functions were built in, see Fig. 10. Each physical signal edge is time stamped with a 1 μs resolution internal timer, and the reaction times of all elements are recorded, published via EPICS and archived for later use. The time measurement QA tests can therefore take the value of the internal timers, with not only a reduced effort but also the availability of analyzing trends from the archived data.

## V. EXPERIENCE OF TWO YEARS OF USE

Out of the four treatment areas at PSI, Gantry 1 and Gantry 2 use the original asynchronous, distributed PaSS since the beginning of its operations in 1996 and 2010 respectively. On the event of modifications, the required unit testing is performed sequentially using an older test stand. The unit testing of each area requires two weeks of work and only tested sequentially parts of the logic at once. The recently installed Gantry 3 uses a new monolithic synchronous PaSS hardware and the newly created test stand. On 2018 Optis was upgraded to the same PaSS technology that was developed for Gantry 3.

On the older technology writing the unit tests document for one area required three weeks and consisted of a textual description and graphically specified waveform timing diagrams. On the new technology, also approximately three weeks are required per area to create the unit test document with the tests described in a formal language. However, the execution time of the tests was reduced to four minutes per area thanks to the automation provided by the new test stand.

In Gantry 3, the development of the new test stand was strongly coupled to the one of the PaSS itself and it permitted testing the hardware before many gantry systems were even available. Since the first release, the Gantry 3 PaSS has been updated five times. The release effort has been reduced to the eventual modification of the affected unit tests, automatic execution at the test stand and a subset of the unit testing, which add up to one to two days of work.

In our experience with the upgrade of Optis PaSS, eight design iterations were required before the final release. The project ran for twelve months and beam time availability at the facility was scarce, so good unit testing was crucial for the success of the upgrade. Also a fast, automated execution of the unit tests allowed for freeing up manpower and reducing the overall length of the project. We are convinced that an automated test stand for unit testing has been a significant improvement in safety level and development time. We will continue to deploy it to the rest of the facility. We can therefore recommend this approach to any facility with similar needs, as the initial development effort is fast offset by the gains.

## VI. CONCLUSIONS

At PSI, a test stand has been developed to automate most of the development QA of the Patient Safety System (PaSS) of our newly installed Gantry 3. The test stand is a PXI based, fully automated computer that executes unit tests under controlled conditions in the laboratory. It is fast, precise and extendable. The unit tests are written in a formal language that was developed based on VHDL for this purpose and which guarantees a compact, easy to read and unambiguous description. A test report is generated automatically upon execution of all test cases. The new safety system itself has also included some self-testing functionality that performs part of the periodical facility QA tests. By automating the unit testing of PaSS, an increased level of safety has been achieved, allowing very complete tests scenarios in less time, therefore freeing up beam time for patient treatment and research. The development cycles in upgrades and bug fixing have also been shortened, as showed in the implementation of this new technology rolled out in the Optis area.Fig. **8**

## APPENDIX

For completeness, a full sample unit test file is provided, showing the inherited VHDL syntax and the nonstandard keywords in bold:

```
DefineMacro SET_INITIAL_CONDITIONS
    -- PaSS Signal Inputs from control system
        TDS_RDY              <= OK;
        START_TREATMENT      <= NOK;
    -- PaSS Signal Inputs from Operating Box
        MODE                 <= NOK;
        OPERATOR_RDY         <= NOK;
EndMacro
...

-- Define test
TestID 1_1_CHECK_BASIC_INTERLOCK
constant t_PAUSE              : time := 100 us;
constant t_VALIDATION_CHECK   : time := 200 us;

Begin
Process Stimuli
 Loop
  Tag Experiment
   callMacro SET_INITIAL_CONDITIONS
    MODE  <= OK; -- Overwrite macro default
  EndTag
  Tag Therapy
   callMacro SET_INITIAL_CONDITIONS
  EndTag
 EndLoop
 Loop
  Tag Master
   AREA_IS_MASTER  <= OK;
  EndTag
  Tag NotMaster
   AREA_IS_MASTER  <= NOK;
  EndTag
 EndLoop
 callMacro DO_PASS_RESET
 Wait for t_PAUSE;
 START_TREATMENT  <= OK;
EndProcess

Process Verification_BASIC_INTERLOCK
 Wait for t_VALIDATION_CHECK;
 Loop
  Tag Experiment
   Assert OUTPUT_ILK = NOK report "No treatment allowed in experiment" severity ERROR;
  EndTag
  Tag Therapy
   Assert OUTPUT_ILK = OK report "Unexpected interlock" severity WARNING;
  EndTag
 EndLoop
 Stop after 100 us;
EndProcess

-- Timing measurements
process Measure_times
 callMacro WAIT_FOR_MEASURE_DELAY
 measure rising_edge(RESET) to falling_edge(OUTPUT_ILK) name "time to clear interlocks";
 measure rising_edge(START_TREATMENT) to falling_edge(OUTPUT_ILK) name "interlock reaction";
EndProcess
EndTestID
```

The previous snippet will unfold into four different tests due to the two loops defined with all combinations of the PaSS in therapy and experiment modes, with and without area mastership. The two tests with the "experiment" tag will fail if an interlock is not triggered when trying to start a treatment and pass otherwise. The opposite will happen with the two tests with the "therapy" tag.